\documentclass{article}
\usepackage{spconf,amsmath,graphicx}
\usepackage{tabularx}
\usepackage[ruled,linesnumbered]{algorithm2e}
\usepackage[labelfont=bf]{caption} %
\usepackage{hyperref,url}
\usepackage{multirow}
\usepackage{booktabs}
\usepackage{cite} 
\usepackage{amssymb} 

\ninept

\renewenvironment{thebibliography}[1]{
  \begin{oldthebibliography}{#1}
  \setlength{\itemsep}{0.1em}
  \setlength{\parskip}{0.1em}
}
{
  \end{oldthebibliography}
}

\usepackage{caption} 
\makeatletter
\renewcommand\section{\@startsection {section}{1}{\z@}%
                                  {-2.5ex \@plus -1ex \@minus -.2ex}%
                                  {1.75ex \@plus.2ex}%
                                  {\normalfont\Large\bfseries}}
\renewcommand\subsection{\@startsection{subsection}{2}{\z@}%
                                     {-1.5ex\@plus -1ex \@minus -.2ex}%
                                     {1.ex \@plus .2ex}%
                                     {\normalfont\large\bfseries}}
\renewcommand\subsubsection{\@startsection{subsubsection}{3}{\z@}%
                                     {-3.25ex\@plus -1ex \@minus -.2ex}%
                                     {1.5ex \@plus .2ex}%
                                     {\normalfont\normalsize\bfseries}}
\renewcommand\paragraph{\@startsection{paragraph}{4}{\z@}%
                                    {3.25ex \@plus1ex \@minus.2ex}%
                                    {-1em}%
                                    {\normalfont\normalsize\bfseries}}
\renewcommand\subparagraph{\@startsection{subparagraph}{5}{\parindent}%
                                    {3.25ex \@plus1ex \@minus .2ex}%
                                    {-1em}%
                                    {\normalfont\normalsize\bfseries}}
\makeatother

\title{Dual-Path Modeling for Long Recording Speech Separation in Meetings}
\name{\begin{tabular}{c}
\it Chenda~Li$^1$, Zhuo~Chen$^2$, Yi~Luo$^3$, Cong~Han$^3$, Tianyan~Zhou$^2$,  \\
\it Keisuke~Kinoshita$^4$, Marc~Delcroix$^4$, Shinji~Watanabe$^5$,  Yanmin~Qian$^1$
\end{tabular}
}
\address{
    $^1$MoE Key Lab of Artificial Intelligence, AI Institute, SpeechLab, Shanghai Jiao Tong University,\\
    $^2$Microsoft Corporation,
    $^3$Columbia University,
    $^4$NTT Corporation,
    $^5$Johns Hopkins University
}
\begin{document}
\maketitle
\begin{abstract}
The continuous speech separation (CSS) is a task to separate the speech sources from a long, partially overlapped recording, which involves a varying number of speakers.
A straightforward extension of conventional utterance-level speech separation to the CSS task is to segment the long recording with a size-fixed window and process each window separately.
Though effective, this extension fails to model the long dependency in speech and thus leads to sub-optimum performance.
The recent proposed dual-path modeling could be a remedy to this problem, thanks to its capability in jointly modeling the cross-window dependency and the local-window processing.
In this work, we further extend the dual-path modeling framework for CSS task. 
A transformer-based dual-path system is proposed, which integrates transform layers for global modeling.
The proposed models are applied to LibriCSS, a real recorded multi-talk dataset, and consistent WER reduction can be observed in the ASR evaluation for separated speech. 
Also, a dual-path transformer equipped with convolutional layers is proposed.
It significantly reduces the computation amount by $30\%$ with better WER evaluation. 
Furthermore, the online processing dual-path models are investigated, which shows $10\%$ relative WER reduction compared to the baseline.
\end{abstract}

\section{Introduction}
\label{sec:intro}

In recent years, the performance of speech separation has been significantly advanced \cite{hershey2016deep,isik2016single,yu2017permutation,kolbaek2017multitalker,chen2017deep,luo2018speaker,luo2017deep,wang2018alternative,luo2018tasnet,luo2019conv,xu2019time,wang2019speech,zeghidour2020wavesplit,luo2020end}. However,
when applied to real-world processing, most existing multi-talker automatic speech recognition (ASR) \cite{yu2017recognizing,settle2018end,chang2019mimo,zhang2020improving,von2020multi,kanda2020investigation} and speech separation systems suffer from two kinds of mismatches. 
First, those systems are usually trained with well-segmented short recordings (e.g. WSJ0-2mix \cite{hershey2016deep}), but in the real world, the duration of conversations varies and could be very long in  scenarios such as meetings.
Second, these systems often assume that the speech is fully overlapped during training, which barely happens in real-world conversations. E.g. as \cite{ccetin2006analysis} suggests, the overlap ratio is usually lower than $30\%$  in a meeting scenario.

The continuous speech separation (CSS) \cite{yoshioka2019css,chen2020continuous} is recently proposed to address the long recording separation for real-world applications, where the long recording is split into smaller length-fixed windows.
The window-level speech separation is performed independently. 
The outputs from adjacent windows are concatenated, or \textit{stitched}, into long output streams.
Ideally, each output stream should only contain overlap-free speech.
And then speaker diarization and ASR can be performed on the overlap-free speech without changing their assumption on single active speaker.
When the window size becomes smaller, given the overlapping characteristics of the real speech, it is reasonable to assume that each window does not contain more than $2$ or $3$ speakers. Thus, the speech separation system trained with short speech and a small number of overlapping speakers can be applied to the long recording speech separation.
One limitation in CSS lies in its incapability of capturing information from long span recording. As each window is processed independently, the receptive field of the separation system is the window length. 
As the context in the long sequence signal usually contains information such as speaker identity, which has been shown beneficial for separation \cite{wang2019voicefilter,delcroix2018single}, a cross-window modeling could potentially further improve the separation performance. 

The recently proposed dual-path (DP) recurrent neural network (DPRNN) \cite{luo2020dual} has been shown promising for speech separation tasks.
The DPRNN splits the long input sequence into smaller, length-fixed windows, and applies two types of RNN layers, namely intra-window RNN and inter-window RNN iteratively on segmented windows. The alternating modeling architecture allows the network to access information across windows that are far apart in time, while maintaining the separation performance for each local window, thus making DPRNN a promising choice for long sequence modeling.
In a recent work \cite{kinoshita2020multi}, the authors applied the dual-path (or \textit{multi-path} ) to long recording separation and achieved promising results. However, the initial experiments only considered a maximal number of $2$ speakers in the entire meeting which only consists of close talk utterances, and the recording-level permutation is aligned across all the windows during training.
In \cite{li2021dualpath}, DPRNN for long recording separation has been initially investigated under a simulated setup.

\begin{figure*}[htb]

  \centering
  \centerline{\includegraphics[width=0.9\linewidth]{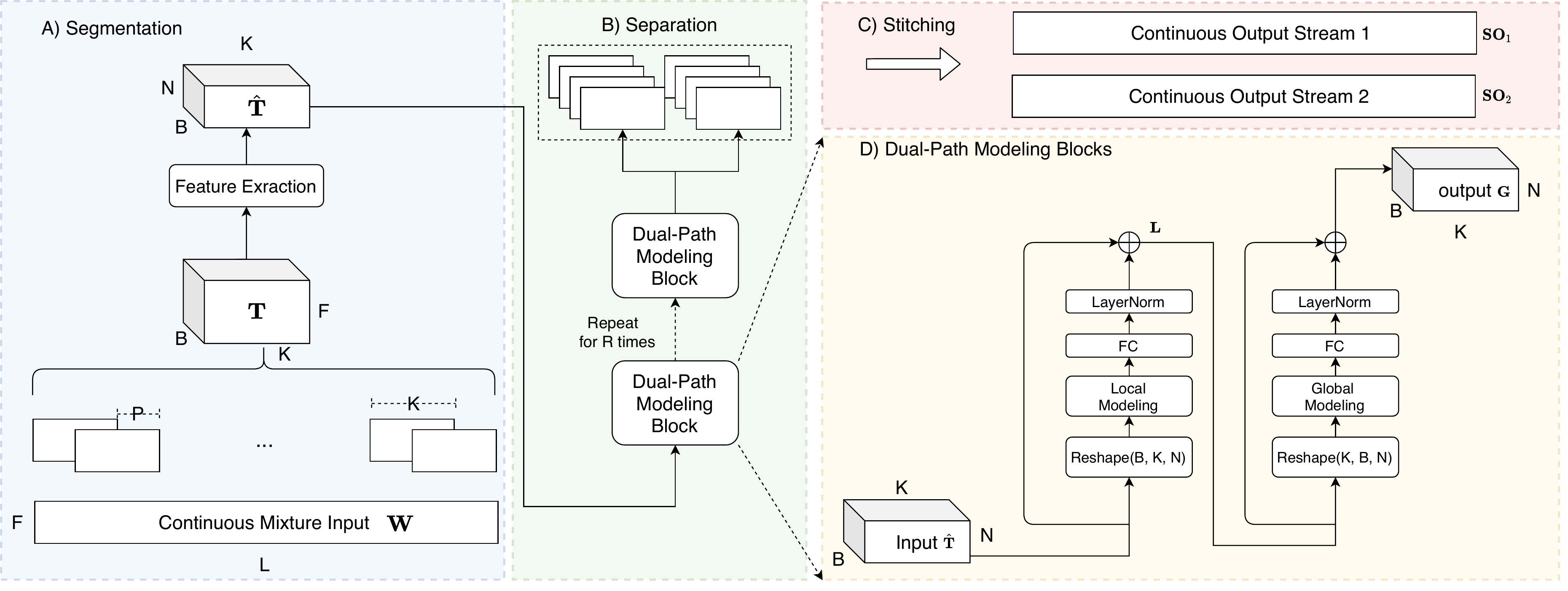}}

\caption{ \textbf{A-C}): The continuous speech separation pipeline. \textbf{A}):The \textit{segmentation} stage splits the long recording into short windows with window size $K$ and hop length $P$. \textbf{B}): The \textit{separation} stage performers the speech separation for each window. \textbf{C}): The \textit{stitching} stage concatenates the separated windows into continuous outputs which only contain non-overlapped speech. \textbf{D}): An illustration of the DP block. 
}
\label{fig:css_pipeline}
\vspace{-0.3cm}
\end{figure*}

In this paper, we further investigate the dual-path modeling in the CSS framework under the realistic setup. 
Similar to DPRNN, we iteratively stack the local and global processing models for long sequence modeling.
We compare two kinds of the most popular models for the dual-path modeling, the RNN and transformer \cite{vaswani2017attention,chen2020dual}.
In the RNN-based DP models, we compare the dual-path bidirectional long-short memory (DP-BLSTM) with the baseline BSLTM on different window sizes.
And the unidirectional LSTM is also explored for global modeling, which allows the system to be deployed to the online meeting processing. 
In the transformer-based DP models, an additional sampling method is proposed to reduce the computation cost as well as improve the separation performance.
The experiments show that the dual-path modeling method not only improves the speech separation performance on simulated testing set, but also effectively reduces the word error rate in automatic speech recognition evaluation on real meeting recordings.

\section{CSS: Task definition and baseline}

The pipeline of conventional continuous speech separation (CSS) is illustrated in Figure \ref{fig:css_pipeline}.
It consists of three stages: \textit{segmentation}, \textit{separation} and \textit{stitching}.

Denote $\mathbf{W} \in \mathbb{R}^{L \times F}$ as the magnitude spectrum of the single-channel continuous mixture input, where $F$ is the number of frequency bins and $L$ is the number of frames.
The \textit{segmentation} stage splits $\mathbf{W}$ into $B$ windows $\mathbf{D}_b \in \mathbb{R} ^{K \times F }, b = 1, \cdots B $ with window size $K$ and hop size $P$.
Then the segmented entire meeting can be presented as a three-D tensor $\mathbf{T} = [\mathbf{D_1} , \cdots , \mathbf{D_B}] \in \mathbb{R}^{B \times K \times F} $, on top of which, a feature extraction module is applied to form the feature for separation step, which has the shape $\hat{\mathbf{T}} \in \mathbb{R}^{B \times K \times N}$ with $N$ referring to feature dimension.

Then for each window,  $C$ streams of output $\mathbf{O}_b \in \mathbb{R}^{K \times F \times C}$ are estimated by the \textit{separation} module, where $C$ is the number of the output channels.
We set $C$ as $2$ in this work, by assuming that the number of overlapped speakers is less than $3$ at most time \cite{ccetin2006analysis}.
The mask based BLSTM separation network is used as the baseline in this work, with phase sensitive mask\cite{erdogan2015phase} as network output.

After obtaining separation result for each window, the stitching step is applied to align the permutation between adjacent window outputs, by finding the permutation that maximizes the similarity from separation results on the shared region between 
adjacent windows. And final result is estimated by a simple overlap-and-add step to connect the local separation result to form output $\mathbf{SO}_{c} \in \mathbb{R}^{L \times F}$ with the same length as mixed signal.

\section{Dual-Path Modeling for CSS}

\subsection{Dual-Path Modeling}

As Figure \ref{fig:css_pipeline}.~B) shows, the DP model stacks $R$ repeats of the basic DP blocks, the details of one DP block is illustrated in Figure \ref{fig:css_pipeline}.~D).
Each DP block consists of two sequence modeling layers, namely the local and global processing layer, where the former focuses on the short term signal modeling, and the latter captures the long span information across windows. With the 3-D tensor as the input feature, global and local layer perform sequence modeling on different axes. By alternating them in a deep DP network, the information from the long sequence can pass across the window, i.e. enabling the network to optimize for the entire long sequence, rather than each local window as in baseline system. Meanwhile, as each sequence layer only models part of the entire sequence, the learning efficiency is significantly improved compared with a single sequence layer for long sequence modeling.

 Denote the bottleneck input feature as $\hat{\mathbf{T}} = [\hat{\mathbf{D}}_1 , \cdots , \hat{\mathbf{D}}_B] \in  \mathbb{R}^{B \times K \times N}$,
the local layer firstly performs the intra-window processing for each individual window $\hat{\mathbf{D}}_{b} \in \mathbb{R}^{K \times N} $:
\begin{align}
	\mathbf{E}_{b}&=f_{\mathsf{local}}\left(\hat{\mathbf{D}}_{b}\right)
\end{align}
Where $f_{\mathsf{local}}(\cdot)$ is the local layer transformation function, $\mathbf{E}_{b} \in \mathbb{R}^{K \times H}$ refers to the processed feature and $H$ is the hidden dimension of the sequential model.
$\mathbf{E}_{b}$ is then processed by a bottleneck fully connect (FC) layer and a layer-norm (LN) \cite{ba2016layer} to build the residual connection\cite{he2016identity}:
\begin{equation}
	\mathbf{L}_{b} = \hat{\mathbf{D}}_{b} + \operatorname{LN}(\operatorname{FC}(\mathbf{E}_{b}))
\end{equation} 
Where $\mathbf{L}_{b} \in \mathbb{R}^{K \times N} $ is the final output of the local processing.
All outputs from all the windows form another 3-D tensor $\mathbf{L} = [\mathbf{L}_1, \cdots, \mathbf{L}_B] \in \mathbb{R}^{B \times K \times N}$. Then, before the global processing, the 3-D tensor is reshaped and indexed as $\mathbf{L}_{k} = \mathbf{L}[:, k, :] \in \mathbb{R}^{B \times N}, k = 1, \cdots, K$.
The global modeling is applied to $\mathbf{L}_{k}$ along the dimension $B$:
\begin{align}
	\mathbf{Q}_{k}&=f_{\mathsf{global}}\left(\mathbf{L}_{k}\right)
\end{align}
where $f_{\mathsf{global}}(\cdot)$ is the global sequential modeling function, and $\mathbf{Q}_{k} \in \mathbb{R}^{B \times H}$ is the global processed feature.
Similar to the local processing, the bottleneck FC, layer-norm and residual connection is applied:
\begin{equation}
	\mathbf{G}_{k} = \mathbf{L}_{k} + \operatorname{LN}(\operatorname{FC}(\mathbf{Q}_{k}))
\end{equation} 
Where $\mathbf{G}_{k} \in \mathbb{R}^{B \times N}$ is the output of the global processing.
The rearranged output  $\mathbf{G} = [\mathbf{G}_1, \cdots, \mathbf{G}_K] \in \mathbb{R}^{B \times K \times N}$ serves as the input of the next DP block.
The output of last DP block $\hat{\mathbf{G}} \in \mathbb{R}^{B \times K \times N}$ is passed to a FC layer with ReLU activation function to generate two T-F masks $\mathbf{M}_b^{1},\mathbf{M}_b^{2} \in \mathbb{R}^{K \times F} $ for each window's magnitude spectrum $\mathbf{D}_{b}$. The masks are applied to the magnitude spectrum by element-wise production to obtain the predicted spectrum $\mathbf{S}_b^{1},\mathbf{S}_b^{2} \in \mathbb{R}^{K \times F} $ for each window.

The window-level permutation invariant training (PIT) is applied during training. It should be noted that the permutation between different windows can be different. 
The training objective is the signal-to-noise ratio (SNR) in the time domain:
\begin{equation}
\mathrm{SNR}(\mathbf{s}, \hat{\mathbf{s}})=10 \log _{10} \frac{\|\hat{\mathbf{s}}\|^{2}}{\|\hat{\mathbf{s}}-\mathbf{s}\|^{2}}
\end{equation}
where $\mathbf{s}$ and $\hat{\mathbf{s}}$ is the estimated and the reference signal of a single window.
The \textit{stitching} is performed during the inference phase.
We calculated the similarity between the predicted mask of adjacent windows to determine the permutation of stitching. 

\begin{figure}[htb]

  \centering
  \centerline{\includegraphics[width=0.9\linewidth]{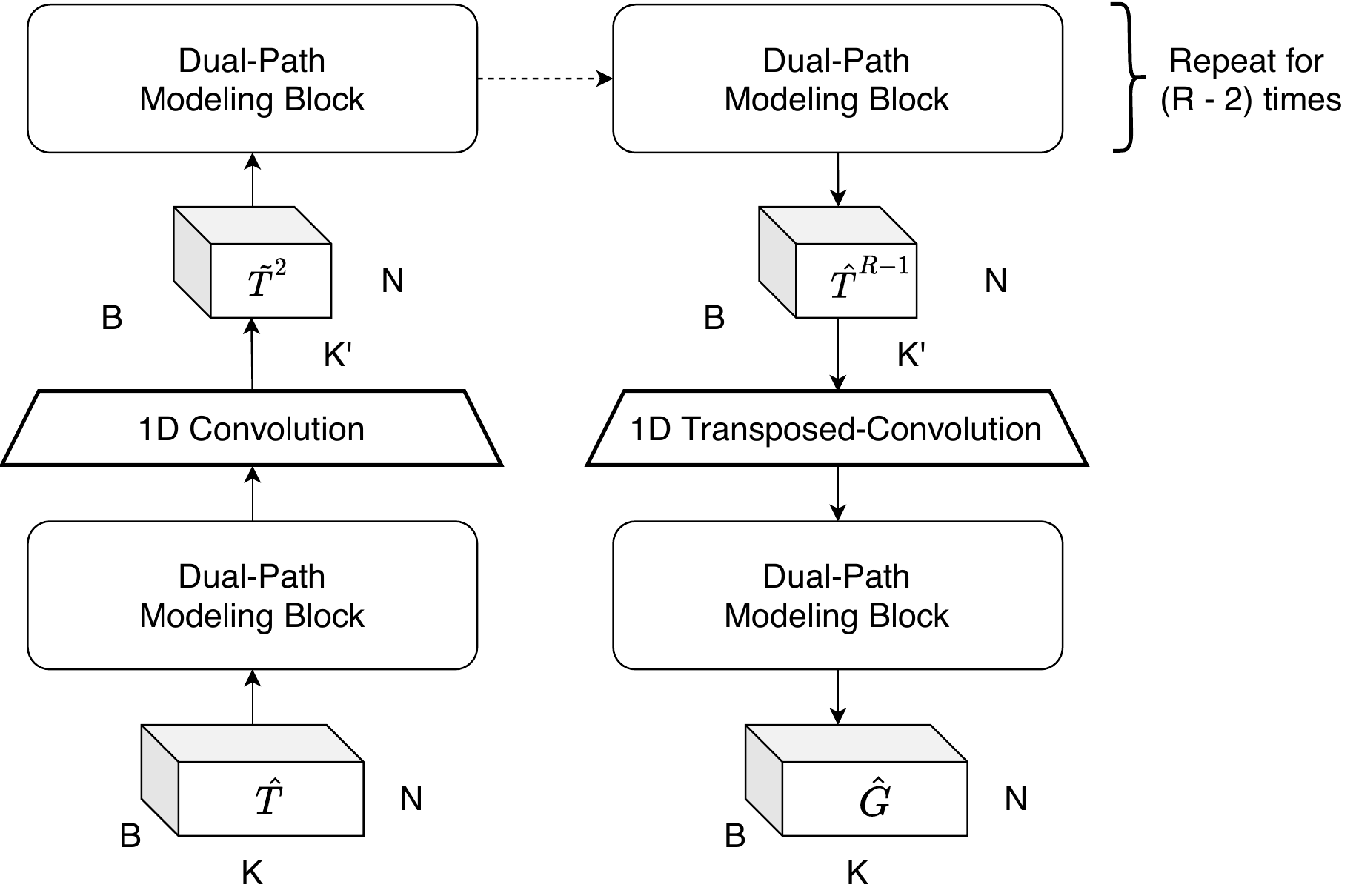}}

\caption{{The boosted dual-path modeling approach. The 1D convolution layer downsamples the feature on the dimension $K$. The size-reduced feature is then processed by the following DP blocks. Before the last DP block, a transposed 1-D convolution layer upsamples the feature to the original length.}}
\label{fig:boosted}
\end{figure}
\subsection{The Boosted Dual-Path Modeling}
\label{sec:boosted_dp}
In this paper, we introduce two updates to the plain DP models, to obtain better separation performance as well as computational efficiency.

First, the transformer encoder layer \cite{vaswani2017attention} is used to replace the RNN in the DP models, which has been shown more effective than RNN in many speech related tasks \cite{karita2019comparative}.
It is noted that a very recent work \cite{chen2020dual} makes the similar update to DPRNN, but the initial experiments are limited in conventional close-talk utterance-level separation. 

Second, we proposed a simple method to improve the DP transformer.
As Figure \ref{fig:boosted} shows, a 1D convolution layer is inserted between the first and the second DP blocks in the separation net.
The 1D convolution is performed on the dimension $K$ and it downsamples the intermediate feature $\hat{T}^{2} \in \mathbb{R}^{B \times K \times N}$ into smaller size $\tilde{T}^{2} \in \mathbb{R}^{B \times K^{'} \times N}$, where $K = \lambda K^{'}$ and $\lambda$ is the sampling factor.
Before the last DP block, the intermediate feature $\tilde{T}^{R-1} \in \mathbb{R}^{B \times K^{'} \times N}$ is processed by a transposed 1D convolution and upsampled back to the tensor  $\hat{T}^{R-1} \in \mathbb{R}^{B \times K \times N}$ which has the same shape as the input, where $R$ is the number of repeated DP blocks.
There are two motivations for this convolution-based resampling in the DP model. 
First, it can effectively reduce the computation cost especially when $R$ becomes large and a proper $\lambda$ is chosen.
Second, the convolution kernel makes the local information better presented in a single frame of one local window, which may benefit the global information interaction.

\section{Experiments}

\subsection{Dataset}

We aim to compare the separation performance in the real application. LibriCSS \cite{chen2020continuous} is used as the testing set. It contains $10$ hours of audio recordings in regular meeting rooms. Each \textit{mini-session}\footnote{Readers can refer to \cite{chen2020continuous} to get more details.} in LibriCSS include $8$ speakers, and the overlap ratio ranges from $0$ to $40\%$.
The recordings are firstly processed by the separation models, and then the continuous input ASR evaluation is conducted.

Given that LibriCSS only contains evaluation data, to train the separation models, we create a training set that consists of artificially simulated noisy and reverberant long-duration audios, based on 16kHz LibriSpeech \cite{panayotov2015librispeech}. The reverbrant speech is created by convolving the clean utterance with the simulated room impulse response(RIR) using image method\cite{allen1979image}. To simulate the long conversation, we create virtual room, each containing multiple RIRs corresponding to different speakers.
We generate $3000$, $300$, and $300$ virtual rooms for training, validation, and testing, respectively.
The width and length of all rooms are randomly sampled between $2$ and $12$ meters, and the height is between $2.5$ to $4.5$ meters.
A microphone is randomly placed within the $2 \times 2$ m$^2$ area in the center of the room, and the height of the microphone is randomly sampled between $0.4$ and $1.2$ meters.
In each simulated room, we randomly choose $10$ candidate speakers from the LibriSpeech \cite{panayotov2015librispeech}.
The locations of these speakers are randomly set at least $0.5$ meters away from the wall, and the height of the speech source is between $1$ and $2$ meters.
The reverberation time is randomly chosen between $0.1$ and $0.5$ seconds.
We simulated $10$ meetings for training in each simulated room.
While generating each meeting, we randomly pick $3-5$ speakers in the current simulated room, and several utterances of these speakers are randomly picked to create the speech mixture.
The duration of the simulated meetings is between $90$ and $100$ seconds.
The overlap ratio of each meeting is uniformly sampled between $50\%$ and $80\%$. 
The overlap region contains up to $2$ speaker, given that more than $2$ speakers talking simultaneously is very rare in real meetings\cite{ccetin2006analysis}.
An additional Gaussian noise with a random SNR from $0$ to $20$ dB is then added to the mixture.
We have totally simulated $30$k, $300$, and $300$ meetings for training, validation, and testing, respectively.

\subsection{Model Configurations and Training Details}

In the feature extraction, the size of short-time Fourier transformation (STFT) is $512$-point and the hop length is $256$.
The window size $K$ in the \textit{segmentation} stage is selected from $\{50, 100, 150, 200\}$, which corresponding to $\{0.8, 1.6, 2.4, 3.2\}$ seconds, respectively.
According to section \ref{sec:intro}, we reasonably assume that each small window only contains up to 2 speakers, and the separation model generates two outputs for each window.
For all the models, the bottleneck feature dimension $N$ is set to $256$.
The RNN-based baseline model is a $4$-layer BSLTM model; each layer contains $512$ forward and $512$ backward hidden units.
The RNN-based DP model contains $2$ repeats of the DP blocks; each block contains $2$ single-layer BSLTMs for local and global processing. The hidden unit is the same as the RNN baseline. Thus the parameter size of the entire model is the same as the baseline.
In the RNN-based DP models, online implementation has also been compared. 
In the online model, the global processing RNN is a unidirectional LSTM with $512$ hidden units.
The transformer baseline contains $10$ transformer encode layers; the attention dimension is $256$, and $4$-head multi-head attention is used. 
The feed-forward layer in the transformer is $1024$ dimensional.
We use $5$ DP blocks in the transformer models to keep the amount of parameter comparable with the baseline.
The Adam optimizer \cite{kingma2014adam} is used in both kinds of models.
The initial learning rates for RNN- and transformer-based models are $0.001$ and $0.002$, respectively.
The warm-up scheduler \cite{vaswani2017attention} is used in the transformer-based models, with $25000$ warm-up steps.
In the RNN-based model training, the learning rate is reduced by $0.9$ every epoch when the validation loss does not decrease.
The batch size is set to $8$.
All the models are trained for $100$ epochs, and the best model on the validation set is chosen.
For the transformer-based models, the parameters of $10$-best models are averaged to get the model for evaluation.
The experiments are conducted using the ESPNet-SE \cite{li2021espnet} toolkit.
\subsection{Window-level Evaluation on Simulation Data}

\begin{table}[th]

	\caption{\footnotesize{Pre-stitching window-level SNR (dB) (2.4s) with different overlap ratios for different models.}}
	\label{tab:segment_level_snr}
	\centering
  \resizebox{\columnwidth}{!}{
   \setlength{\tabcolsep}{1.5mm}{

    \begin{tabular}{l|c|ccccc}
    \toprule
    \multirow{2}{*}{Models} & \multirow{2}{*}{\begin{tabular}[c]{@{}c@{}}Model \\Size (M)\end{tabular}} &  \multicolumn{5}{c}{Overlap ratio in  \%  } \\ 
     &    & 0 & 0-25 & 25-50 & 50-75 & 75-100  \\ 
    \hline
    {BLSTM}  &  13.9  & 16.25 & \textbf{7.92}  & 9.42 & 9.19  & 8.60   \\
    {DP-BLSTM}  & 13.9  & \textbf{16.38} & 7.83  & \textbf{9.91} & \textbf{9.69}  & \textbf{8.87}   \\
    \midrule
    Trans.  & 8.2   & 16.15 & 8.15  & 9.79 & 9.49  & 8.79   \\
    DP-Trans.  &  8.2  & \textbf{16.21} & 8.03  & 9.85 & \textbf{9.61}  & \textbf{8.91}   \\    DP-Trans. +   &  10.1  & 16.14 & \textbf{8.17 } & \textbf{9.87} & 9.49  & 8.67   \\

    \bottomrule
    \end{tabular}
    }}

\end{table}

We firstly evaluated the window-level SNR before the \textit{stitching} stage on the simulation test set.
The results are listed in Table.\ref{tab:segment_level_snr}.
The SNR scores are reported on different overlap ratios.
Results in Table.\ref{tab:segment_level_snr} show that the DP models can consistently beat their baseline with comparable parameter size except for the overlap ratio $0-25 \%$ conditions.

\subsection{Continuous ASR Evaluation on LibriCSS}
\begin{table}[th]
  \caption{{ WER (\%) evaluation on LibriCSS for continuous speech separation with different models. All our models in the table use the window size of $2.4$s. 0S/L \cite{chen2020continuous}: 0\% overlap ratio with short/long silence.} }
  \label{tab:asr_evaluation_1}
  \centering
  \resizebox{\columnwidth}{!}{
   \setlength{\tabcolsep}{1.5mm}{
  \begin{tabular}{ l c c | c c  c c c c }
    \toprule
    \multirow{2}{*}{Systems} &  \multirow{2}{*}{\begin{tabular}[c]{@{}c@{}}\footnotesize{Model} \\ \footnotesize{Size (M)}\end{tabular}} & \multirow{2}{*}{\begin{tabular}[c]{@{}c@{}}MACs \\(Giga)\end{tabular}} & \multicolumn{6}{c}{Overlap ratio in \%}  \\
     &  & &  0S & 0L & 10 & 20 & 30 & 40 \\
    \midrule
    Mixture \cite{chen2020continuous} & - & - & 15.4 & 11.5 & 21.7 & 27.0 & 34.3 & 40.5  \\
    \footnotesize{BLSTM}\cite{chen2020continuous} & & & 17.6 & 16.3 & 20.9 & 26.1 & 32.6 & 36.1 \\
    \midrule
    \footnotesize{BLSTM} & 13.9 & 54.4 & \textbf{15.3} & 13.6 & \textbf{18.6} & 24.9 & 30.4 & 33.9 \\
    \footnotesize{DP-BLSTM} & 13.9 & 54.4 & 16.0 & \textbf{12.1} & \textbf{18.6} & \textbf{24.1} & \textbf{29.1} & \textbf{32.7} \\
    \midrule
    Trans & 8.2 & 31.5 & 16.0   & 14.4  & 19.0 & 22.6  & 29.5  & 33.5 \\
    DP-Trans. & 8.2 & 31.5 & 15.6 & 14.7 & 18.8 & 22.8 & \textbf{29.1} &
    \textbf{32.3} \\
    DP-Trans. + & 10.1 & \textbf{21.4} &  \textbf{14.2} & \textbf{12.3} & \textbf{17.4} & \textbf{22.4} & \textbf{29.1} & 32.5 \\
    \bottomrule
  \end{tabular}}
  }

\end{table}

The continuous ASR evaluation follows the same manner in \cite{chen2020continuous}, with the default ASR backend from LibriCSS dataset. 
After the \textit{stitching} stage, the separated overlap-free speech is fed into the pertained ASR evaluation pipeline, the word error rates (WERs) of different models are reported in Table.\ref{tab:asr_evaluation_1}. 

In Table.\ref{tab:asr_evaluation_1}, both of our BLSTM and the transformer baseline are stronger than those reported in  \cite{chen2020continuous} (the 2nd row). 
The DP-BLSTM gets better WERs compared to the BLSTM baseline, except for the 0S results;
The improvement of the DP transformer is relatively smaller, but it still shows effectiveness in the $40\%$ overlapped meetings.
The DP transformer equipped with convolution layers (last row in table) reduces the amount of multiply-accumulate (MAC) operations by $30\%$ relatively.
At the meantime, it also shows better WER on most conditions, especially in meetings with low overlap ratios.

\begin{table}[t]
  \caption{{WER (\%) evaluation on LibriCSS for continuous speech separation with different local processing window size. The comparison is conducted on dual-path and the baseline BLSTMs.}}
  \label{tab:online}
  \centering
   \resizebox{\columnwidth}{!}{
  \begin{tabular}{c | c c | c c  c c c c }
    \toprule
    \multirow{2}{*}{\begin{tabular}[c]{@{}c@{}}Window \\Size\end{tabular}} & \multirow{2}{*}{\begin{tabular}[c]{@{}c@{}}Dual- \\Path\end{tabular}} & \multirow{2}{*}{\begin{tabular}[c]{@{}c@{}}Window \\Online\end{tabular}} & \multicolumn{6}{c}{Overlap ratio in \%}  \\
      & & & 0S & 0L & 10 & 20 & 30 & 40 \\
    \midrule
    \multirow{3}{*}{0.8s}& No  & Yes & 16.1 & 12.7 & 19.9 & 25.0 & 31.8 & 36.4 \\
     					& Yes & No & 15.0 & 12.8 & 18.1 & 22.9 & 28.3 & 31.7 \\
     					 & Yes & Yes & 14.7 & 13.2 & 18.6 & 24.3 & 29.3 & 32.7 \\

    \midrule
    \multirow{3}{*}{1.6s}& No  & Yes & 16.2 & 14.5 & 20.1 & 25.1 & 31.3 & 34.6 \\
     					& Yes & No & 15.0 & 12.0 & 18.4 & 23.0 & 28.6 & 31.6 \\
     					 & Yes & Yes &15.8 & 12.9 & 18.5 & 23.6 & 29.9 & 32.9 \\

    \midrule
     \multirow{3}{*}{2.4s} & No & Yes &15.3 & 13.6 & 18.6 & 24.9 & 30.4 & 33.9 \\
           						& Yes & No &16.0 & 12.1 & 18.6 & 24.1 & 29.1 & 32.7 \\
      						& Yes & Yes & 15.6 & 12.4 & 18.4 & 23.6 & 29.9 & 32.8 \\
    \midrule
    \multirow{3}{*}{3.2s} & No & Yes &15.5 & 13.4 &  19.4 &  24.7 & 30.7  & 33.7  \\
      						& Yes & No & 15.2 & 12.3 & 18.7 & 24.3 & 29.9 & 33.7 \\
      					 & Yes & Yes & 15.9 & 12.7 & 18.8 & 23.8 & 29.9 & 33.4 \\

    \bottomrule
  \end{tabular}
  }

\end{table}

\subsection{Comparison Window Lengths and Online Processing}

The bidirectional modeling (BLSTM or self-attention) is used for the cross-window information interaction, so the above DP models can not be directly applied to the online processing.
One straightforward way to enable the online processing for the DP models is to replace the BLSTM with uni-directional LSTM(uni-LSTM) for the cross-window processing. It is also possible to build the DP transformer for online processing, but we leave it for future work. Note that under the LSTM global modeling setup, the maximum system latency is equal to the local window size. 
Table \ref{tab:online} compares the online DPRNNs with the offline DPRNNs and the baseline BLSTM.
The models with different window sizes have been compared. 
Results in Table \ref{tab:online} show that, for the local baseline model, the WERs get worse when the window size becomes smaller.
It is because the smaller window size leads to less local information. 
While, for the DP models, they always outperform their baseline local models.
One interesting finding is that, the smaller window size achieves better WERs for the DP models. 
One possible explanation for this is that the smaller window size splits more windows, leading to finer resolution for global modeling and thus enhancing the information pass across windows.
The last rows in each section of Table \ref{tab:online} list the WERs of the online dual-path models, which also show their efficacy compared to the baseline.

\section{Conclusion}

In this paper, we investigated the dual-path modeling for long recording speech separation in real meeting scenarios. We explored both the RNN- and Transformer-based dual-path models, and the experimental results showed that the dual-path models outperformed the baseline consistently in the CSS task. We proposed a dual-path transformer with convolutional sampling, which reduces the computation amount by $30\%$, and get $3\%$ relative WER reduction on LibriCSS meeting recordings compared to the baseline. The online dual-path model also achieved $10\%$ relative WER reduction, which makes it a strong candidate for online continuous speech separation.

\section{Acknowledgments}
Chenda Li and Yanmin Qian were supported by the China NSFC projects (No. 62071288 and U1736202).
The work reported here was started at JSALT 2020 at JHU, with support from Microsoft, Amazon, and Google.

\bibliographystyle{IEEEtran}
\bibliography{refs}

\end{document}